\documentclass[pra,twocolumn, aps,amssymb,nofootinbib,longbibliography,showpacs,superscriptaddress]{revtex4-1}

\usepackage{graphicx}
\usepackage{dcolumn}
\usepackage[justification=RaggedRight]{caption}
\usepackage{siunitx}
\usepackage{hyperref}
\usepackage{svg}
\usepackage{float}
\usepackage{physics}
\usepackage{color}

\def\CT{{\cal T}}
\def\up{\uparrow}
\def\down{\downarrow}

\begin{document}

\title{Adiabatic ground state preparation in an expanding lattice}
\author{Christopher T. Olund}
\affiliation{Department of Physics, University of California at Berkeley, Berkeley, CA 94720, USA}
\author{Maxwell Block}
\affiliation{Department of Physics, University of California at Berkeley, Berkeley, CA 94720, USA}
\author{Snir Gazit}
\affiliation{Racah Institute of Physics and The Fritz Haber Research Center for Molecular Dynamics, The Hebrew University of Jerusalem, Jerusalem 91904, Israe}
\author{John McGreevy}
\affiliation{Department of Physics, University of California at San Diego, La Jolla, CA 92093, USA}
\author{Norman Y. Yao}
\affiliation{Department of Physics, University of California at Berkeley, Berkeley, CA 94720, USA}
\affiliation{Materials Science Division, Lawrence Berkeley National Laboratory, Berkeley, California 94720, USA}

\date{\today}

\begin{abstract}
We implement and characterize a numerical algorithm inspired by the $s$-source framework [Phys. Rev.~B 93, 045127 (2016)]
for building a quantum many-body ground state wavefunction on a lattice of size $2L$ by applying adiabatic evolution to the corresponding ground state at size $L$, along with $L$ interleaved ancillae. The procedure can in principle be iterated to repeatedly double the size of the system. We implement the algorithm for several one dimensional spin model Hamiltonians, and find that the construction works particularly well when the gap is large and, interestingly, at scale invariant critical points. We explain this feature as a natural consequence of the lattice expansion procedure. This behavior holds for both the integrable transverse-field Ising model and non-integrable variations. We also develop an analytic perturbative understanding of the errors deep in either phase of the transverse field Ising model, and suggest how the circuit could be modified to parametrically reduce errors. In addition to sharpening our perspective on entanglement renormalization in 1D, the algorithm could also potentially be used to build states experimentally, enabling the realization of certain long-range correlated states with low depth quantum circuits.

\end{abstract}

\maketitle

\section{Introduction}

A deep lesson of late-20$^\text{th}$-century physics is the renormalization group (RG) philosophy: many body physics is organized scale-by-scale.
The fruits of this lesson have been assimilated well into our understanding of classical statistical physics and of perturbative quantum field theory \cite{RevModPhys.71.S358, wilson_renormalization_1974}.
In strongly-correlated quantum systems, however, we still have a great deal to learn, in particular about eigenstates and even groundstates of local model Hamiltonians.

Most of many-body Hilbert space is fictional, at least in the sense that it cannot be reached from a product state by time evolution with local Hamiltonians in a time polynomial in system size \cite{Poulin:2011zz}.
Ground states of local Hamiltonians are even more special: generically (with few exceptions arising from an overabundance of gapless excitations) the entanglement entropy of large-enough subregions satisfies an area law \cite{2008PhRvL.100g0502W}.
This statement is supported by a great deal of evidence, and has been rigorously proved for gapped systems in 1D \cite{2007JSMTE..08...24H}.
%

Importantly, much of the area-law corner of Hilbert space can be efficiently parameterized using tensor networks.
This has been done with several different tensor network geometries, such as matrix product states (MPS)~\cite{PhysRevB.55.2164, McCulloch_2007,szasz2018observation,PhysRevLett.122.040603} in 1D, and projected entangled pair states (PEPS)~\cite{jordan2008classical,PhysRevB.84.165139,PhysRevB.99.165121,PhysRevA.77.052306} and isometric tensor networks~\cite{zaletel2019isometric, soejima2019isometric} in 2D.
These parameterizations have proven to be very effective variational ansatzes in a wide range of circumstances \cite{Orus:2013kga, PhysRevB.94.155123} \footnote{We note that there are also \textit{non-variational} algorithms for finding ground states that make use of tensor networks, some of which are provably efficient in some circumstances \cite{landau-vazirani,PhysRevB.73.094423, arad_rigorous_2017, roberts_implementation_2017}}. 
In particular, the density matrix renormalization group (DMRG) algorithm can be understood as a variational optimization on the MPS manifold \cite{PhysRevLett.69.2863, schollwock2011density}.
Despite their successes, variational algorithms based on area-law tensor network ansatzes face some limitations.
Specifically, in gapless phases, or at critical points, entanglement entropy can diverge with subsystem size making these area-law tensor networks sub-optimal variational manifolds.
It is also known that there exist even area law states that do not have an efficient MPS representation~\cite{PhysRevLett.100.030504}.
Finally, many tensor networks are difficult to efficiently optimize in $D>1$ \cite{PhysRevB.94.195143, PhysRevB.94.155123}.

Developing numerical methods for gapless phases and critical points requires understanding a richer entanglement structure than area-law states exhibit -- we must account for the amount of entanglement at each length scale.
The process of organizing our understanding of the entanglement in a quantum state scale-by-scale is sometimes called {\it entanglement renormalization} \cite{PhysRevLett.99.220405,Vidal:2008zz}.
So far, the best-developed implementation of this idea is the multiscale entanglement renormalization ansatz (MERA), which is a state-of-the-art variational ansatz for the study of 1D quantum critical points \cite{Pfeifer:2008jt, Evenbly2013,PhysRevB.79.144108,PhysRevLett.100.240603,PhysRevB.99.241105}.
MERA has also inspired several variants such as deep MERA (DMERA)~\cite{kim2017robust} and an analytic construction continuous MERA (cMERA)~\cite{PhysRevLett.110.100402, PhysRevD.99.085005}.

Despite the successes of MERA, developing a deeper understanding of entanglement renormalization remains a key challenge in condensed matter physics.
More generally, existing tensor network methods leave room for improvement in several ways.
First, the numerical values of the optimal tensors found in this way are difficult to interpret or directly relate to analytic results; the procedure is essentially a black box.
Second, and more practically, the variational minimization of the expectation value of the Hamiltonian requires sweeping across the lattice many times, an often-costly procedure which has many opportunities to get stuck in locally-optimal configurations.

In this paper, we introduce and benchmark a numerical algorithm for entanglement renormalization that takes small steps towards alleviating some of these issues.
In particular, we provide a numerical implementation of the so-called $s$-source framework, originally introduced in Ref.~\cite{Swingle:2014qpa}.
We note that the purpose of this work is to implement $s$-source and characterize its accuracy; we leave a rigorous resource analysis to future studies.
%

We now briefly describe the $s$-source formalism; a more thorough explanation is provided in Ref.~\cite{Swingle:2014qpa}.
Let $H_L$ be a hamiltonian defined on a $d$-dimensional lattice of size $L^d$ and $\ket{\psi^L}$ be the associated ground state.
The Hamiltonian family $\{H_L\}$ belongs to an $s$-source fixed point if $\ket{\psi^{2L}}$ can be constructed by applying a quasi-local unitary $U$ to $s$ copies of $\ket{\psi^L}$ and some unentangled ancilla degrees of freedom.
In many cases, we expect the adiabatic theorem to provide a construction of such a quasi-local unitary: if there is a gapped path from $H_L$ to $H_{2L}$ then adiabatic evolution along this path will suffice.
There is evidence that many known states are $s$-source fixed points, including trivial insulators ($s=0$), chiral insulators ($s=1$) and various field theories \cite{Swingle:2014qpa}.
Examples with $s > 1$ are known as well, including fracton models \cite{2014PhRvB..89g5119H, 2017arXiv171205892S, PhysRevB.99.115123}.
Belonging to an $s$-source fixed point constrains the growth of entanglement with system size, and in particular when $s < 2^{d-1}$ implies an area law for the entanglement entropy of subregions \cite{Swingle:2014qpa}.
While the construction in Ref.~\onlinecite{Swingle:2014qpa} is more general, we will specialize our numerical exploration to one dimensional spin chains.

The key advantage of $s$-source is its ability to generate long-range entangled states using a constant depth circuit for $s \geq 1$.
We illustrate this by comparing it to some more intuitive state preparation schemes.
It is well known that building highly entangled states from a product state with local gates requires extensively deep quantum circuits \cite{harrow2018approximate}.
Even with $\ket{\psi^L}$ as a resource, not all renormalization schemes generate long-range entanglement.
In 1D, for example, one might consider concatenating two copies of a ground state end-to-end, and then acting with a local unitary to ``glue" the states together.
Unfortunately, constructing a long-range entangled state in this manner is not possible since the local unitary cannot strongly entangle distant spins in the two halves.
In the $s$-source framework in 1D, we circumvent this issue by intercalating $L$ ancilla spins between the spins that make up $\ket{\psi^L}$, thus expanding the underlying lattice.
Crucially, this implies that a quasilocal unitary only needs to \emph{locally} redistribute the rescaled entanglement structure. We will refer to the state formed by interleaving ancillae and $\ket{\psi^L}$ as the ``$s=1$ input state", or just the ``$s$-source input state" where $s=1$ is to be understood.
In contrast, we would call a product state of $2L$ spins an ``$s=0$ input state."


\begin{figure}[t!]
	\centering
	\includegraphics[width=3in]{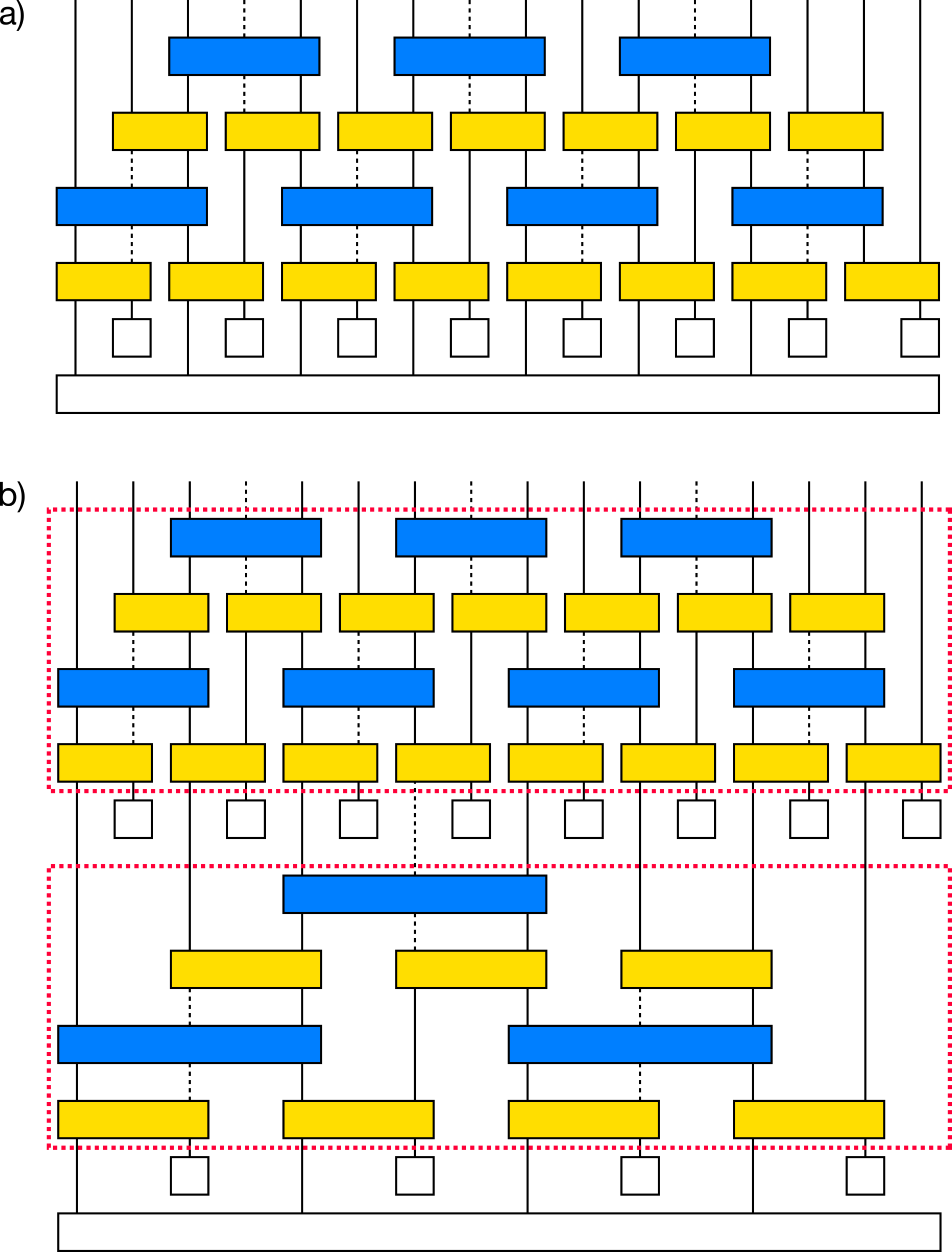}
	\caption{Circuit diagrams for the s-source renormalization procedure. a) A single layer of the circuit which takes an eight spin ground state (the large black box) and eight ancillae (small black boxes) and after applying the turn ``on'' (yellow) and turn ``off'' (blue) unitaries produces an approximation to the sixteen spin ground state. b) A two superlayer circuit which starts from the four spin ground state and produces an approximation to the sixteen spin ground state; each block outlined by a red dashed line represents a superlayer made up of two on and two off layers.}
\label{fig:CircuitDiagrams}
\end{figure}

To adapt the $s$-source construction to a numerical setting, one must decide how to implement the quasi-local unitary.
One possibility would be to perform quasi-adiabatic evolution via an algorithm like time evolving block decimation (TEBD)~\cite{PhysRevLett.91.147902,PhysRevLett.93.040502}.
In this work, we take an alternative route and fix a tensor network structure informed by the analytic Trotterization of the adiabatic evolution, which we refer to as the $s$-source tensor network.
The tensor network we obtain closely resembles a single layer of the MERA tensor network, and repeated application indeed results in a flavor of MERA.
In principle, the Trotterized adiabatic unitary provides an exact expression for the $s$-source tensor network, but explicitly calculating it is generically computationally intractable.
Instead, we seek to \emph{find} the corresponding tensor network through other means.
One approach, which is possible in certain limiting cases, is to make use of the Trotterized structure to determine analytic expressions for the constituent tensors.
More generally, we can determine the $s$-source network variationally by minimizing the expectation value of $H_{2L}$ with $\ket{\psi^{L}}$ as input (Fig.~\ref{fig:CircuitDiagrams}a).
At its core, since we are still using a \emph{variational} approach to identify the $s$-source tensor network, our prior concerns of becoming stuck in a locally-optimal configuration still apply. 

Although we still resort to variational optimization, the existence of the analytic expression defining $s$-source allows us to physically interpret the resulting network and encourages us to optimize it in novel ways.
For example, when considering multilayer $s$-source (red boxes in Fig.~\ref{fig:CircuitDiagrams}b), we think of each layer as an independent adiabatic expansion and optimize it \textit{separately} by minimizing the energy of the appropriate Hamiltonian at that scale, given the input generated by the preceding layers.
%
%
We emphasize that this is slightly different from the usual optimization of MERA, where one sweeps over the entire network multiple times.
In general, our numerical procedure incurs larger errors than optimizing across all layers, and hence is sub-optimal when compared to standard global optimization approaches. 
Even so,  it is  less computationally expensive and, as we will explore below, can still perform well in certain cases of interest. 
In addition, by utilizing a greedy numerical algorithm, our implementation allows us to numerically test the validity of the adiabatic construction at the heart of the $s$-source approach.

We will benchmark our implementation's performance on the standard transverse field Ising model (TFIM)
\begin{equation}
H_{TFIM}=-J\sum_{\langle ij\rangle}\sigma^z_i\sigma^z_j - h\sum_i \sigma^x_i
\end{equation}
as well as the TFIM with integrability-breaking  perturbations~\cite{PhysRevB.95.214410,PhysRevX.7.031011}.
The TFIM has several limits that provide intuition about how $s$-source should behave in general.
The first limit of interest is deep in the ferromagnetic phase ($h<J$), where the ground state at finite size is approximately a symmetric Greenberger--Horne--Zeilinger (GHZ) state, i.e. $(\ket{\up\ldots\up}+\ket{\down\ldots\down})/\sqrt{2}$ (as opposed to the symmetry broken state such as $\ket{\up\ldots\up}$ that one would typically consider in the thermodynamic limit).
Building a GHZ state from a product state with local gates requires a circuit of extensive depth, but with a size $L$ GHZ input the size $2L$ GHZ state can be prepared using a single layer of nearest-neighbor controlled-not gates.
Similarly, the exact finite-size ferromagnetic ground state at finite magnetic field is another simple example of a state with long-range entanglement that cannot be built from a product state with a finite depth circuit.
One would expect  the same to be true of a gapped ground state with nontrivial topological order: while we would have no hope of building such a state with a low depth circuit from a product input, an $s$-source input could allow one to construct a good approximation.

In contrast, the paramagnetic phase ($h>J$) is easier to approximate with a product state input.
Deep in the phase the ground state is almost a product state, and the correlations that do exist are short-range.
Using an $s$-source input naturally doubles the length scale of those correlations, and thus, to build the ground state one has to first \emph{remove} those unwanted correlations before building the desired ones back in; for a product state input, we would only have to do the latter.
Even if the $s$-source constructed state has low error, our effort is wasted; we could have done even better with less work by starting with a product state.

\begin{figure}
	\centering
	\includegraphics[width=3.375in]{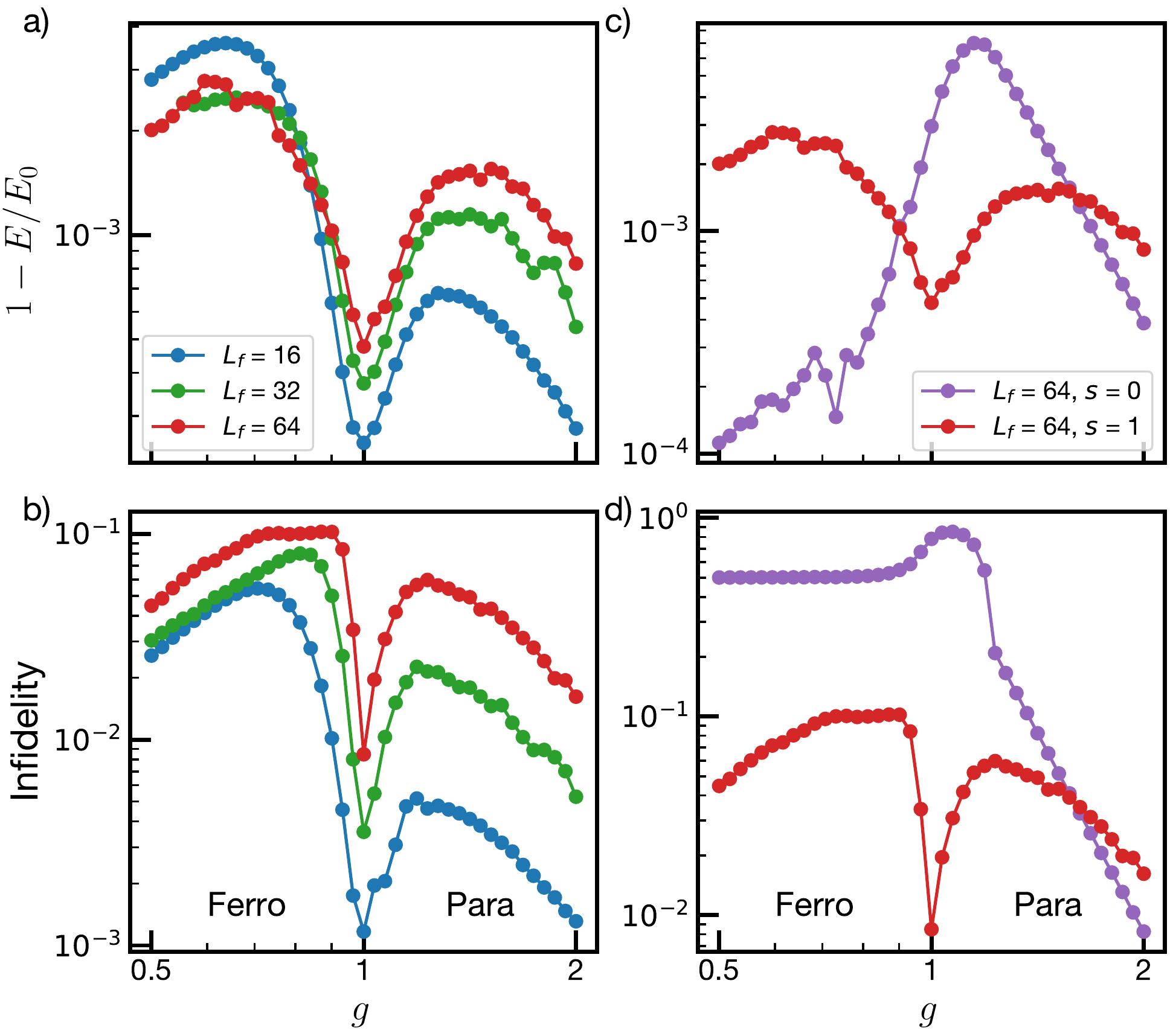}
	\caption{a) Relative energy error and b) infidelity of the TFIM s-source ground state as a function of $g$ for several system sizes. Both go to 0 deep in either phase as one would expect, but there is also a local minimum at the critical point due to the scale invariance of the system. c) Relative energy error and d) infidelity for $L_\text{f}$ = 64 with with either the normal s-source input (s = 1) or product state input (s = 0). Using a product input state gives better energies deep in both phases, but has an infidelity over 0.5 in the ferromagnetic phase as long range entanglement cannot be generated. By either metric, s = 1 input gives substantially smaller errors near the critical point.}
   \label{fig:tfim_errs}
\end{figure}

The most interesting case is at the critical point.
Here, we generically observe a local minimum in the error as a function of the transverse field strength (Fig.~\ref{fig:tfim_errs}).
In some ways this is quite surprising; the existence of the analytic $s$-source construction relies on the adiabatic theorem, which in turn requires a gap.
Of course, there will always be a gap due to finite system size; however, one naively expects that such a small finite-size gap would force one to use a longer adiabatic evolution time, thus incurring larger Trotter errors when approximating the adiabatic unitary with a local circuit.
However, the scale invariance of the TFIM critical point makes it particularly amenable to approximation by $s$-source.
At the critical point, the correlation length of the ground state scales with system size, so when we insert ancillae and hence trivially double the length scale of correlations, we actually achieve the proper long-range entanglement structure.
We then correct the short range details with the local circuit.
We note that the location of the error minimum remains at the critical point even when one adds generic pertubations to the TFIM, consistent with the expectation that this behavior should generalize to other continuous phase transitions.

Our paper is organized as follows.
In Sec.~\ref{sec:algorithm}, we give a  precise description of both the $s$-source algorithm and our numerical implementation.
In Sec.~\ref{sec:results}, we benchmark our numerical implementation by applying it to several standard 1D spin chain models: first, the (integrable) transverse field Ising chain (TFIM), next, the TFIM with a longitudinal field which is non-integrable and has no symmetries, and finally the TFIM with a symmetry-preserving but integrability-breaking term.
In Sec.~\ref{sec:analytics}, we develop some analytic understanding of the circuit in the large-gap limit.
Finally, in Sec.~\ref{sec:discussion}, we summarize our results and discuss potential future directions of study.

\section{s-source Algorithm and Numerical Implementation}

\label{sec:algorithm}

In the $s$-source framework, we regard the entanglement present in the ground state at linear system size $L$ as a resource for constructing the ground state at system size $2L$.
Rather than attempting to directly prepare the macroscopic ground state of a model Hamiltonian, we suppose we are given $s$ copies of the ground state at system size $L$,
and design a circuit which doubles the system size. 
That is, we seek a unitary map which produces the ground state at size $2L$ from $s$ copies of the ground state at size $L$ times a collection of factorized ancillary qubits.
Iterating this doubling procedure yields a circuit which produces the ground state in the thermodynamic limit from $s$ copies of the (easily-determined) ground state of a small cluster of sites.
We note, as previously discussed, that such a size-doubling map can exist even when the state represents a nontrivial phase and cannot be constructed from a product state via a low-depth local unitary circuit. 

Our numerical implementation will focus on $s$-source with $s=1$. 
%
When the Hamiltonian is gapped, one can immediately write down an expression for the $s$-source unitary using the adiabatic theorem. 
Let $\tilde{H}_L$ be the operator that acts as $H_L$ on the odd lattice sites only.
Now, consider a time-dependent Hamiltonian $H(t)$ which interpolates between
\begin{equation}
    H(0) = \tilde{H}_L- \sum_{i\textrm{ even}} X_i 
\end{equation}
and
\begin{equation}
H(T) = H_{2L}.
\end{equation}
Here,  $X_i$ are operators which put the ancillary qubits into a product ground state.
The unitary operator which generates this time evolution is then:
\begin{equation}
U=\CT e^{ -i\int_0^T  H(t)dt}.
\label{eq:adiabaticU}
\end{equation}
Of course, if we could generally compute the full adiabatic unitary explicitly we could also solve the \emph{easier} problem of just finding the exact ground state at system size $2L$! 
We can, however, imagine Trotterizing this unitary to get an approximation built out of \textit{local} unitaries that is tractable enough to make further progress. 
When the gap is large we can find these component local unitaries analytically, as we explore in Sec.~\ref{sec:large-h}. 

For the moment, however, we observe that even without actually doing the time-ordered integral, one can see upon which spins the local unitaries act; the terms in the leading order of the Trotter expansion will act on the same spins as do terms in either $H_L$ or $H_{2L}$. 
Since we will work with nearest-neighbor Hamiltonians, one can think of the terms coming from $H_{2L}$ as turning on interactions between the spins of our original $L$ site system and the ancillae (which are now nearest neighbors after the interleaving step), and we can interpret the terms coming from $H_{L}$ as turning off interactions between the original spins (which are no longer nearest neighbors). 
Keeping these leading order terms, we get an approximate tensor network for $U$ as shown schematically in Fig.~\ref{fig:CircuitDiagrams}a. Although we justify the circuit structure perturbatively, we will see from our numerics that it is still capable of generating approximate ground states even when a perturbative expansion would not converge. 
The order of the layers is in principle arbitrary, although some choices are more computationally efficient than others. 
We also note that one could choose to Trotterize into larger blocks and that doing so would improve the approximation in exchange for increased circuit optimization becoming much more computationally expensive. 
Later, in Sec.~\ref{sec:perturb}, we will see exactly how introducing longer range blocks reduces errors deep in either phase of the transverse field Ising model.

In our numerical implementation, we treat the tensor network as a variational ansatz built out of arbitrary unitaries. We minimize $\langle H_{2L}\rangle$ over those component unitaries to get an approximation for $|\psi^{2L}\rangle$.
As the reader may have noted, the circuit that we end up obtaining is, in fact, a MERA, albeit one with a particular circuit structure and where we have cut off some number of layers at the smallest scale. However, we are thinking of this MERA as being ``upside down"; rather than starting with a large state and repeatedly coarse graining, we start with a small state and scale up.

There is a fundamental tension between making the adiabatic evolution time $T$ larger to reduce adiabaticity errors and making $T$ smaller to reduce Trotterization errors for a fixed depth circuit. This tension disappears in the extreme limit of a large gap wherein we can determine U analytically, as we will describe in Sec.~\ref{sec:large-h}.



We now describe the actual circuit ansatz used, and explain how we numerically optimize it to find an approximate ground state. 
%
Suppose we have a solution for the ground state of $H$ for an $L$ spin system $|\psi^L\rangle$ in matrix product state (MPS) form.
We construct the $2L$ spin input state $|\phi^{2L}\rangle$ by identifying spin $i$ ($1\leq i\leq L$) of the $L$ particle system with spin $2i-1$ of the $2L$ spin system, and then placing ancillary spins on the remaining sites. We note that the orientation of these ancillae does not matter as any single spin rotation can be absorbed into the circuit. 
%
Next, we construct a quantum circuit described by a total unitary $U_{T}$. We build this circuit in four layers: (i) applying two spin unitaries $U_A^i$ to each pair of spins $(2i-1,2i)$ for $1\leq i \leq L$, (ii) applying unitaries $U_B^i$ to pairs of spins $(4i-3,4i-1)$ for $1\leq i \leq L/2$, (iii) applying unitaries $U_C^i$ to pairs $(2i+1,2i)$ for $1 \leq i \leq L-1$, and finally (iv) applying $U_D^i$ to pairs $(4i+1,4i-1)$ for $1\leq i \leq L/2-1$.
The unitaries $U_A^i$ and $U_C^i$ correspond to turning on the new couplings between the original $L$ spins and the ancillae, and the  unitaries $U_B^i$ and $U_D^i$ and correspond to turning off the couplings between the original spins. 
A schematic of this setup for $L=8$ can be seen in Fig.~\ref{fig:CircuitDiagrams}a. 
We can also repeat this procedure multiple times, successively inserting ancillae and then applying four layers of the circuit (which we will henceforth call a ``superlayer") to repeatedly double the size of the input state. 
As an example, a two-superlayer circuit is illustrated in Fig.~\ref{fig:CircuitDiagrams}b.

In order to numerically optimize the circuit, we minimize the energy $E=\langle \widetilde{\psi}^{2L}|H_{2L}|\widetilde{\psi}^{2L}\rangle$ where $|\widetilde{\psi}^{2L}\rangle=U_T|\phi^{2L}\rangle$. 
In particular, we begin with an initial circuit (which could either be a random circuit or an educated guess) and then consider $E$ to be a function of each of the individual local unitaries comprising $U_T$. 
We then sweep over all of these component unitaries multiple times using the conventional MERA update procedure described in \cite{PhysRevB.79.144108}. 
For the interested reader, we provide some additional details about the numerical optimization in Appendix~\ref{A:optimization}.

As mentioned above,  this optimization procedure generally only finds a local minimum of the energy; if one wants to reliably find the global minimum, it is necessary to do this variational search many times with different initial conditions. 
To optimize a multilayer circuit, we pursue a greedy algorithm: for each superlayer, we minimize the expectation value of $H_{2L}$ over the unitaries in the $L$ to $2L$ layer with all preceding layers held fixed. 
The intuition behind this approach is that the adiabatic construction should in principle guarantee the existence of a multilayer circuit such that its first $k$ superlayers generate the  ground state at size $2^k L$. 
To be more precise, if we consider the analytic construction where we have the exact quasilocal adiabatic unitaries at our disposal, we know that one can construct the state $\ket{\psi^{4L}}$ from $\ket{\psi^L}$ by applying the adiabatic unitary $U^{L\rightarrow 2L}$ to get $\ket{\psi^{2L}}$ and then $U^{2L\rightarrow 4L}$ to get $\ket{\psi^{4L}}$. 
This suggests that a greedy approximation of each layer could in principle be effective. Of course, re-optimizing all superlayers at each scale is at least as accurate and in some cases may yield much lower errors. 
However, full circuit optimization comes at a significant computational cost and we find that the greedy approach performs surprisingly well. 
Before presenting our numerical benchmarking results, we wish to emphasize that our approach, following the $s$-source philosophy, attempts only to find an optimal {\em adiabatic} trajectory, namely one which utilizes information from previous layers. In particular, we do {\em not} attempt a global energy minimization, as with standard MERA optimization schemes.

\section{Numerical Benchmarking Results}

\label{sec:results}

To benchmark our numerical implementation we consider three 1D models: the transverse field Ising model (TFIM), a mixed-coupling Ising model (MCIM), and a mixed-field Ising model (MFIM), with Hamiltonians:
\begin{align}
	 H_{TFIM}&=-J\sum_{\langle ij\rangle}\sigma^z_i\sigma^z_j - h\sum_i \sigma^x_i \nonumber \\ 
	 H_{MCIM}&=-\sum_{\langle ij\rangle}\left(J_x\sigma^x_i\sigma^x_j+J_z\sigma^z_i\sigma^z_j\right) - h\sum_i \sigma^x_i \nonumber \\
	 H_{MFIM}&=-J\sum_{\langle ij\rangle}\sigma^z_i\sigma^z_j - \sum_i \left(h_x\sigma^x_i+h_z\sigma^z_i\right).
\end{align}
The TFIM sets our baseline understanding for how $s=1$ $s$-source performs in three limits: a short range correlated unique ground state (the paramagnetic phase), an almost-degenerate long range correlated ground state (the ferromagnetic phase), and at a critical point. 

We quantify our implementation's performance using both the relative error in energy (which we minimize) and the many-body infidelity, i.e.~the overlap mismatch between the $s$-source state obtained at size  $2L$ and the ``exact'' DMRG wavefunction at the same size, $1-|\langle \widetilde{\psi}^{2L}|\psi^{2L}\rangle|^2$. 
As a point of reference, we compare this performance to that of $s=0$ $s$-source, optimizing the same circuit structure with a product state input. 
We also study the consequences of truncating our approximation of the quasi-local unitary to include only nearest-neighbor gates.
Finally, in order to understand the propagation of errors in our numerical $s$-source algorithm, we analyze the performance of multilayer circuits. 
%


\emph{Benchmarking via the TFIM model}---Since the TFIM is integrable, in this case, we calculate energy errors relative to the exact values. For the MCIM and MFIM models, we benchmark against energies obtained via DMRG. 
In addition, we use DMRG to generate our initial input MPS  states for $s$-source for all three models (restricting to a specific $\mathbb{Z}_2$ parity sector when appropriate).
In Fig.~\ref{fig:tfim_errs}a, we plot the relative error in energy for a single layer of $s$-source for the TFIM as a function of $g=h/J$ for several values of the input system size $L_0$. 
To be specific, this means that we start with the ground state at $L_0$ and perform a single layer of our $s$-source algorithm to obtain an approximate ground state at $L_{\textrm{f}} = 2 L_0$, whose energy we then compare with the exact value. 
Similarly, Fig.~\ref{fig:tfim_errs}b depicts the many-body infidelity, which exhibits the same qualitative behavior.
In all of our numerics, we ensure that the input state in the ferromagnetic ($g<1$) phase is the non-symmetry broken ground state. 

As one expects, the error decreases deep in either the ferromagnetic or paramagnetic phase.
Indeed, because the gap is large in these regions, there must exist a suitable $s$-source adiabatic unitary that minimizes both non-adiabatic and Trotter errors.
Less expected, from this adiabatic perspective, is the existence of a local minimum in the error at the TFIM's critical point, $g=1$, despite the fact that the gap vanishes at this point. 
Naively, one might have expected that this would lead to an error maximum instead. 
In fact, this is exactly what does happen if we start with a product state input ($s=0$) instead of the $s=1$ $s$-source input state, as can be seen in Fig.~\ref{fig:tfim_errs}c. 
%

To understand this $s=1$ local minimum, we note that the correlation length diverges at the critical point and it is impossible to capture these correlations starting from a product state and using a low-depth local quantum circuit. 
However, if we start with the size $L$ ground state (as we do in $s=1$ $s$-source), then correlations of length $L$ become correlations of length $2L$ upon  ancillae insertion. 
In principle, at the critical point, this is exactly what we desire from the size $2L$ ground state; we emphasize once again that this is precisely the same intuition which underlies MERA and that our circuit is in fact a type of MERA with a ``cut-off'' at small scales. 

To further check this intuition, we can define a single site energy error for the TFIM as 
\begin{equation}
\epsilon(i)=-\frac{J}{2}\left(\sigma^z_{i-1}\sigma^z_i+\sigma^z_i\sigma^z_{i+1}\right)-h\sigma^x_i,
\end{equation}
and then take a Fourier transform to define a momentum-resolved energy error $\epsilon(k)$. Doing this, we found that the momentum resolved error was only significant for momenta of $k=0$, $k=\pi/4$, and $k=\pi/2$ ($k$ in units of inverse lattice spacing). The $k=0$ component is just the total energy error, whereas the $k=\pi/4$ and $k=\pi/2$ components correspond to errors of characteristic length scale 2 and 1 lattice spacings, respectively. These are, of course, exactly the length scales at which the nearest-neighbor and next-nearest-neighbor gates comprising our circuit act. There is no corresponding dip in $\epsilon(k=\pi/4)$ or $\epsilon(k=\pi/2)$ at the critical point, consistent with our understanding that the dip in the overall energy error really is due to ancilla insertion and not the local action of the circuit.

While this built in doubling of input correlations is beneficial at criticality, it can  be detrimental in other regimes. 
This can be seen by comparing the performance of $s=1$ $s$-source with $s=0$ $s$-source deep in the paramagnetic phase ($g>1$ in both Fig.~\ref{fig:tfim_errs}c,d). 
While both errors are scaling toward zero as $g$ increases, the scaling is worse for the $s=1$ input.
Here, the true ground state is  short-range correlated, approaching a product state for large $g$.
Thus, constructing the size $2L$ ground state with an $s=1$ input actually involves first getting \emph{rid} of all the doubled correlations.
%

Looking only at energy errors (Fig.~\ref{fig:tfim_errs}c), the above statement would also appear to apply  deep in the ferromagnetic phase ($g<1$). 
However, the many-body infidelity tells a different story. 
In particular, although the energy errors for $s=0$ $s$-source scale better than $s=1$ $s$-source, the fidelity does not (Fig.~\ref{fig:tfim_errs}d).
To understand this behavior, we note that at $g=0$, the ground state manifold of the TFIM is two-fold degenerate, consisting of the symmetric and anti-symmetric cat states. 
For finite but small $g$, these states will be split in energy by an exponentially small gap $\sim g^L$. 
Until $g$ is nearly one, any linear combination of these two states will have approximately the same energy, and an ``all up" like combination can be constructed from a product state input to give a low energy error. 
However, it is impossible to construct a cat state from a product state using a circuit with sub-extensive depth. 

As a result, the infidelity of the  $s=0$ state is always greater than $0.5$ throughout the  entire ferromagnetic phase, as the zeroth order piece of the true ground state cannot be constructed. In contrast, with an $s=1$ input state, a size $L$ cat state can be  used to create a size $2L$ cat state by using controlled NOT gates between each  pair of  original and ancilla spins. 
We expect that this behavior should generalize to certain classes of topological states.
In particular, because one cannot change a topological invariant by acting with local unitaries, it is impossible to build such states from a product state input. On the other hand, using an $s=1$ input preserves the topological character of the state.

\begin{figure}
	\centering
	\includegraphics[width=3.375in]{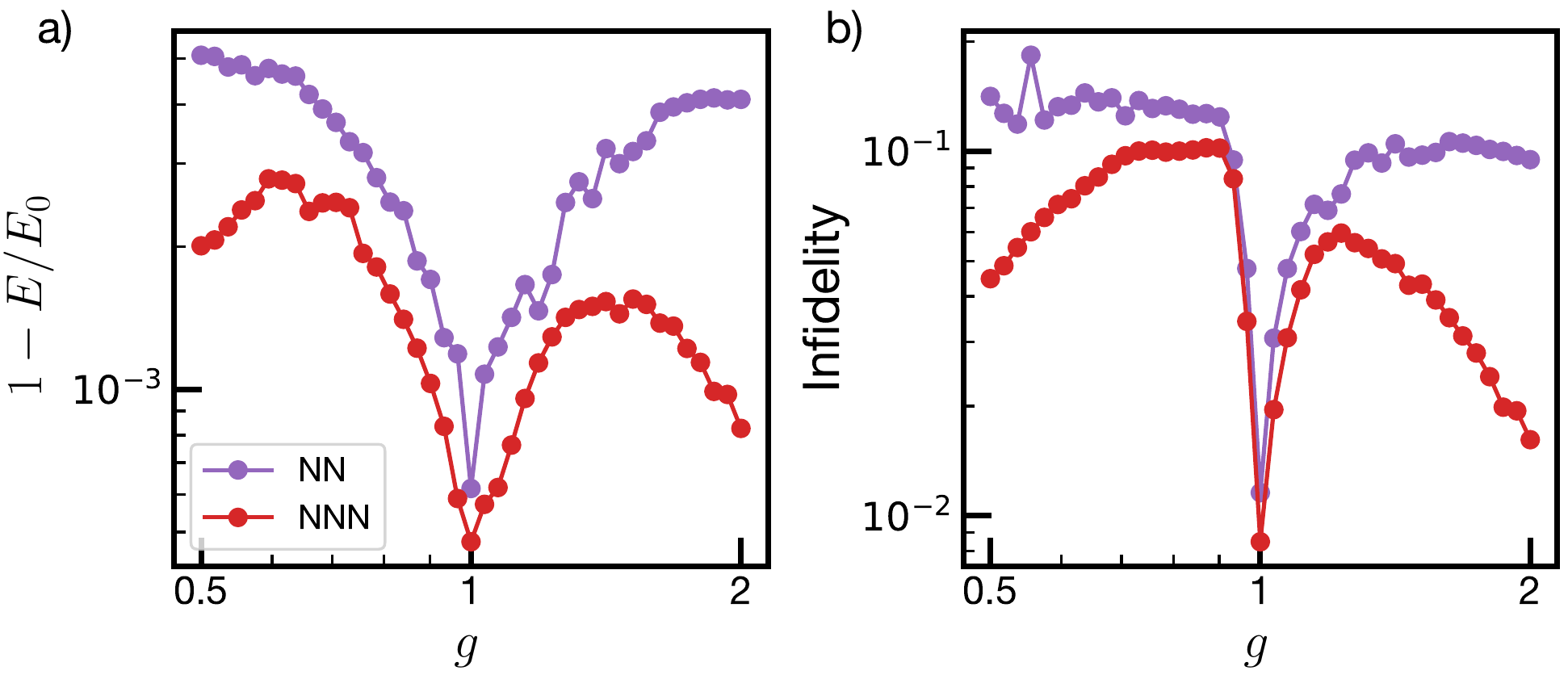}
	\caption{a) Relative energy error and b) infidelity of the TFIM for both the ``standard" $s$-source circuit with both nearest-neighbor (NN) and next-nearest-neighbor (NNN) unitaries and a simplified circuit with only NN unitaries, both for $L_f=64$. The general shape of the error curves, notably including the local minimum at the critical point, are similar for both circuits. The NN circuit does almost as well as the NNN circuit at the critical point, but the errors fall off more slowly than for the NNN circuit deep in either phase.}
   \label{fig:tfim_bin}
\end{figure}

Next, we turn to studying the effect of changing the range of the quasi-local unitary approximation by restricting our circuit to include \emph{only} nearest-neighbor gates.
A comparison of the resultant energy errors and infidelities are shown in Fig.~\ref{fig:tfim_bin}. 
The nearest-neighbor circuit still exhibits a local error minimum at the critical point, and in fact nearly achieves the accuracy of the longer ranged $s$-source circuit there. 
This implies that the long-range correlations built in via ancillae insertion and the ability to perform nearest-neighbor corrective gates are the most important features for accurately constructing a state at the critical point.  %
Moving away from the critical point, one sees the advantage of the next-nearest-neighbor circuit geometry; we will show in Sec.~\ref{sec:analytics} that the range of individual gates in the circuit determines the scaling of the error with $g$ deep in either phase.

\begin{figure*}
	\centering
	\includegraphics[width=0.8\textwidth]{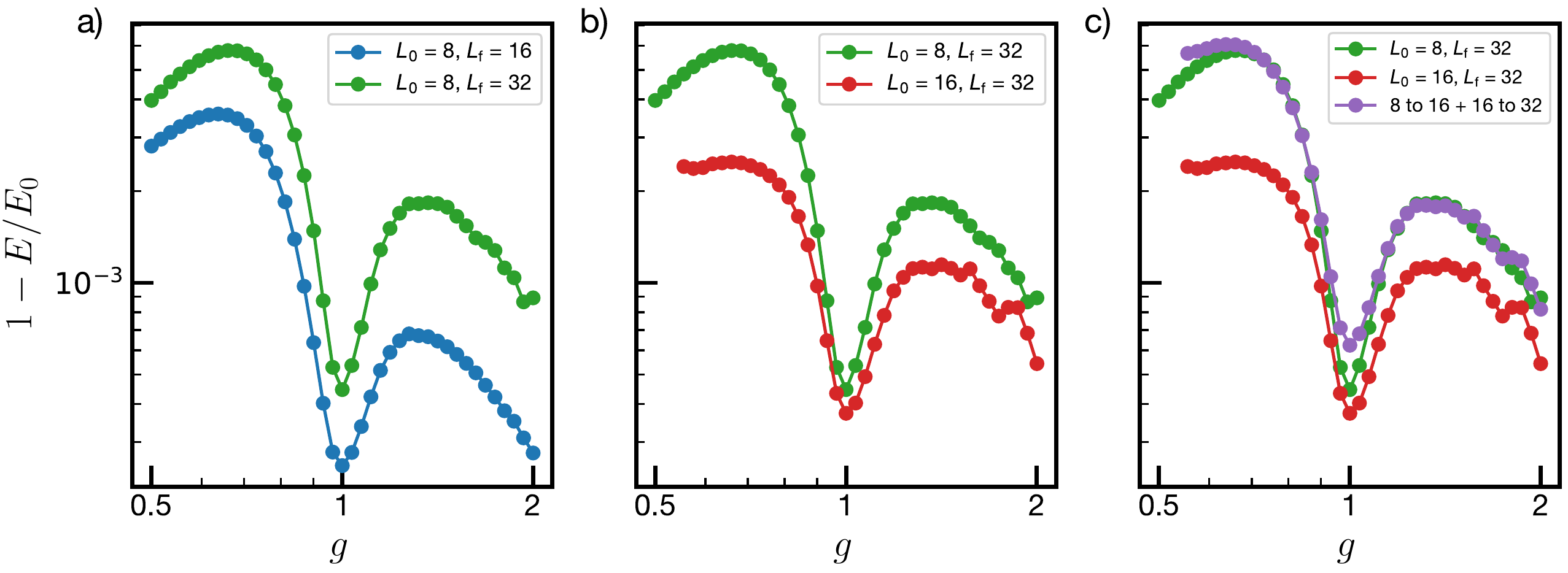}
	\caption{Relative energy error in the TFIM for multilayer circuits. a) Error comparison starting with $L_0$ = 8 for 1 and 2 layer circuits. b) Error comparison ending with $L_f$ = 32 for 1 and 2 layer circuits. Error compounds reasonably with successive layers, and in particular is not much worse than single layer optimization at the critical point. c) We hypothesize that the multilayer error is subadditive, as illustrated here. Except where we have failed to find the global minimum, the $L_0$ = 8 to $L_f$ = 32 error is bounded by the  $L_0$ = 8 to $L_f$ = 16 error plus the  $L_0$ = 16 to $L_f$ = 32 error, with the multilayer circuit substantially outperforming this bound at the critical point.}
   \label{fig:tfim_multi}
\end{figure*}

\emph{Multilayer $s$-source}---Our preceding discussion focuses on single-layer $s$-source, where one starts with a size $L_0$ input state and ends with a size $L_f = 2L_0$ final state.
In multilayer $s$-source, we start from a size $L_0$ state and perform the $s$-source construction $n$ times to get an approximate size $L_f=2^nL_0$ state; we use the approximate state from one superlayer as  the input state for the next. 
As aforementioned, in our numerics, we take a ``greedy'' approach where we optimize each superlayer in isolation rather than sweeping back and forth.
In principle simultaneously optimizing superlayers should improve accuracy, but it would come at a substantial computational cost. 

Fig.~\ref{fig:tfim_multi} depicts the energy errors for multilayer $s$-source for the  TFIM as a function of both $L_0$ and  $L_f$.
Although it is difficult to make sharp statements, it appears that errors are not accumulating, per layer, super-linearly.
We hypothesize that the multilayer error obeys the following bound: $\mathcal{E}_{L\rightarrow 4L}\le \mathcal{E}_{L\rightarrow 2L}+\mathcal{E}_{2L\rightarrow 4L}$, where $\mathcal{E}_{L_0\rightarrow L_f}$ is the relative energy error for the optimal $L_0$ to $L_f$ $s$-source state. 
The analogous statement for accumulated infidelities holds trivially (if one were to optimize the $s$-source circuit by minimizing infidelity instead of energy).
In Fig.~\ref{fig:tfim_multi}c, one can see that this proposed  bound appears to hold for a two superlayer circuit.

\begin{figure}
	\centering
	\includegraphics[width=2.4in]{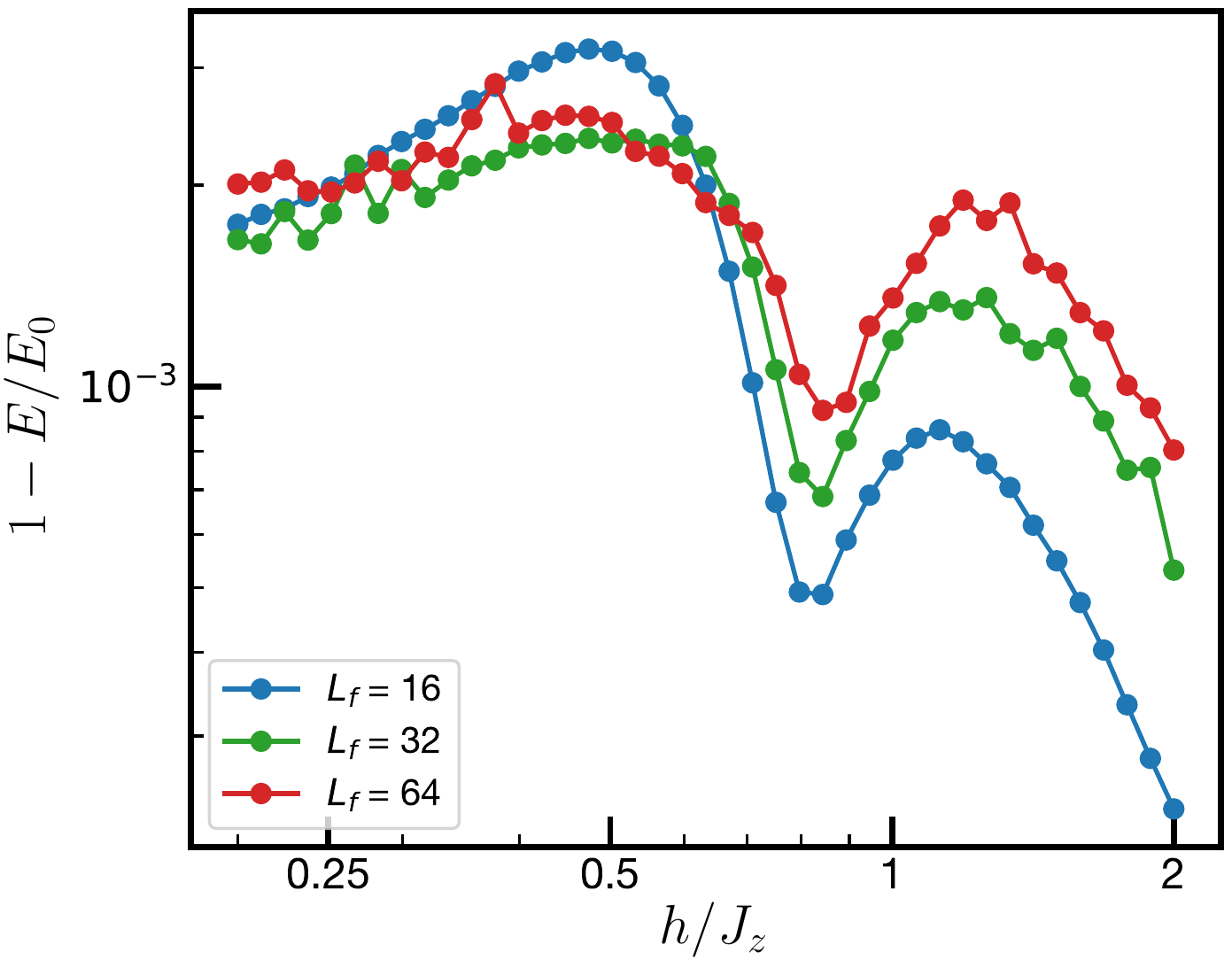}
	\caption{Relative energy error for the mixed coupling Ising model with $J_x = 0.1J_z$. The features of the error curve are qualitatively similar to the transverse field case, and the local minimum remains at the model\text{'}s critical point.}
   \label{fig:xx_errs}
\end{figure}

\begin{figure}
	\centering
	\includegraphics[width=2.4in]{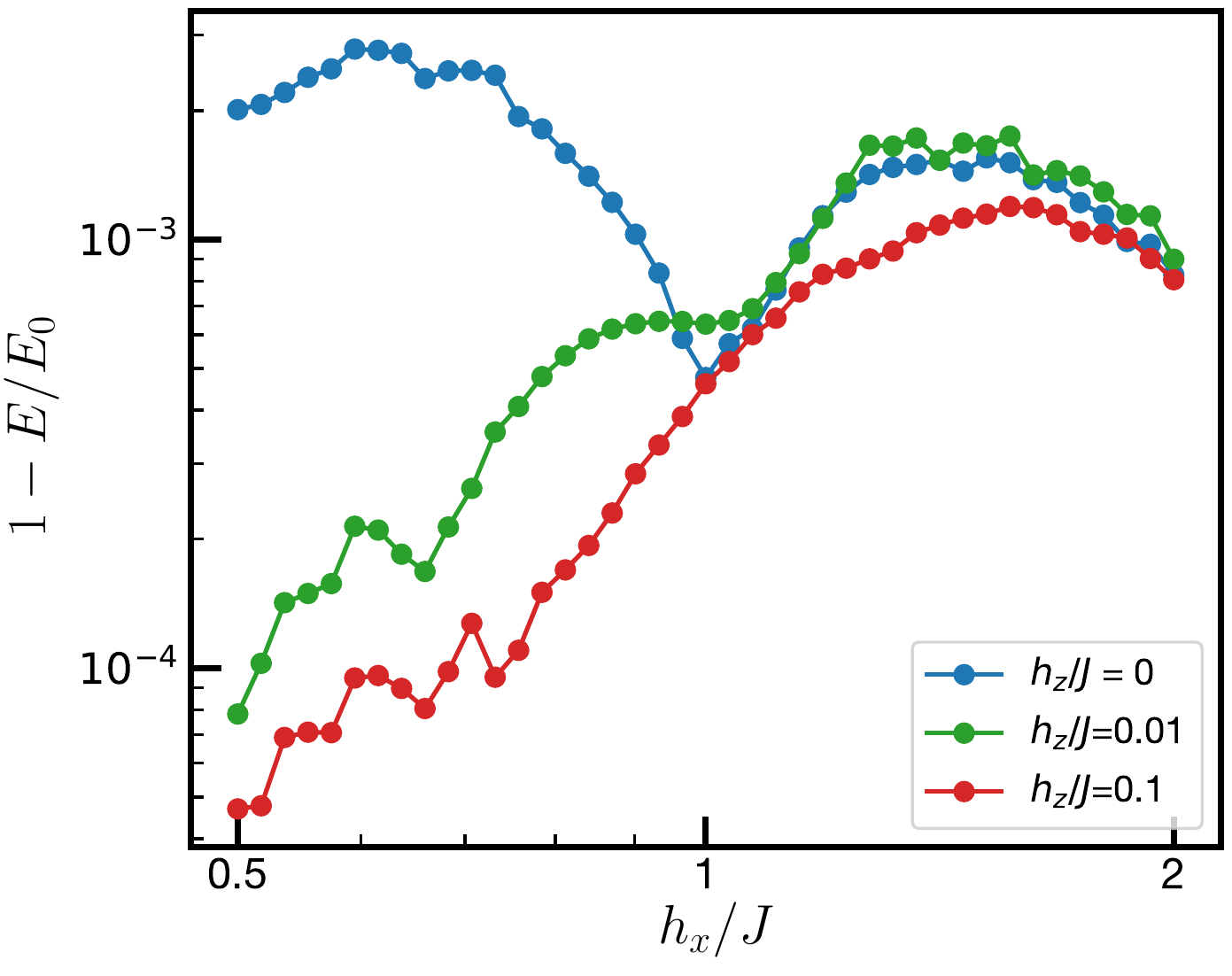}
	\caption{Relative energy error for the mixed field Ising model for $L_f$ = 64. For sufficiently large $h_z$, there is no longer a local minimum in the error as there is no scale invariant point.}
   \label{fig:z_errs}
\end{figure}

\emph{Non-integrable models}---Finally, we now turn to applying the single-layer $s$-source algorithm to the non-integrable MCIM and MFIM models. 
In Fig.~\ref{fig:xx_errs}, we begin by showing the energy errors as a function of input size for the MCIM model. 
The qualitative features of the error curve are analogous to what we have already discussed in the TFIM case; the error decreases  deep in either phase, and there is still a local minimum in the error at the phase transition. 
This minimum is consistent with our prior expectations since the  critical point of the MCIM is still described by a conformal field theory.
In Fig.~\ref{fig:z_errs}, we plot the energy errors for a number of different longitudinal field  strengths for the MFIM model. As $h_z$ is increased, the local minimum flattens out and then vanishes, consistent with the lack of a scale invariant point.

\section{Analytic analysis of errors}
\label{sec:analytics}
\subsection{Analytic tensors in the large field limit}
\label{sec:large-h}
Our effort to express the adiabatic $s$-source unitary as a local, low-depth circuit faces two competing constraints. 
In order to have a good approximation, we need to be able to use both the adiabatic theorem (which requires that $T^{-1}$ be small compared to the gap) and the Trotter decomposition (which requires that $T^{-2}$ be large compared to commutators between different blocks of the Hamiltonian). 
%
In particular, in the large $h/J$ limit of the TFIM, one can satisfy both of the above constraints. By moving into the interaction picture and expanding the time ordered exponential to leading order, we find (see Appendix \ref{A:large-h} for details) that the nearest-neighbor (``on") unitaries are (Fig.~1), to leading order in $h/J$, given by $U=e^{-iH_\mathrm{eff}}$, where
\begin{equation}
H_\mathrm{eff} =-\frac{J}{8h}(\sigma_z^1\sigma_y^2+\sigma_y^1\sigma_z^2).
\end{equation}
Similarly, the next-nearest-neighbor (``off") unitaries (Fig.~1) are given by $U=e^{iH_\mathrm{eff}}$ with the same $H_\mathrm{eff}$.

One can perform the same calculation in the mixed field/coupling models. Defining $h=\sqrt{h_x^2+h_z^2}$ and $\tan\eta = h_z/h_x$, we obtain an effective Hamiltonian

\begin{equation}
H_\mathrm{eff}= \alpha(\sigma_x^1\sigma_y^2+\sigma_y^1\sigma_x^2) + \beta(\sigma_y^1\sigma_z^2+\sigma_z^1\sigma_y^2).
\end{equation}
where 
\begin{equation}
\begin{split}
\alpha=&\frac{J_x}{32h}\left(7\sin{\eta}+3\sin{3\eta}\right) -\frac{3J_z}{8h}\cos^2\eta\,\sin\eta\\
\beta=&\frac{3J_x}{8h}\cos\eta\,\sin^2\eta-\frac{J_z}{32h}\left(7\cos{\eta}-3\cos{3\eta}\right)
\end{split}
\end{equation}
and again the ``on" and ``off" unitaries are given by $U=e^{-iH_\mathrm{eff}}$ and $U=e^{iH_\mathrm{eff}}$, respectively.

\subsection{Perturbative analysis}
\label{sec:perturb}
In this subsection, we will use a perturbative analysis to explore how the error of the optimal $s$-source circuit varies with our system parameters. 
In our numerics, we are ultimately using a  variational (in energy) method to solve for the circuit; thus, one cannot  analytically calculate the error directly, but in a perturbative regime, we can compute how the leading order correction to the $s$-source wavefunction scales. 
As a bonus, this procedure naturally suggests additional tensors one could include in the circuit to further suppress errors.
Though including such tensors  would come with a computational cost for our numerical implementation, it is possible that they could be more natural for certain experimental geometries where long-range interactions are present \cite{Gadway_2016, DOHERTY20131, monroe2019programmable, Barredo1021, Endres1024}.

Here we present the results of this analysis for the TFIM.
Our basic strategy is to first figure out the exact $s$-source circuit at a fixed point  (i.e. either $h/J=0$ or $J/h=0$), and then to perturb around this fixed point.
In other words, once we factor out the fixed point portion, we will parameterize each unitary in the circuit as $\exp(-iH)=\mathbb{I}-iH+\ldots$ with $\left\Vert H\right\Vert = \mathcal{O}(h/J)$ or $\mathcal{O}(J/h)$. 
We then consider whether there exists such a circuit that will take our initial state to the target state correctly to a given order in  perturbation theory.

For the $J/h=0$ fixed point (with ancillae inserted in the direction of the field), the $s$-source circuit is simply the identity and reproduces the size $2L$ ground state perfectly since both $|\psi^L\rangle$ and $|\psi^{2L}\rangle$ are product states. 
Deep in the paramagnetic phase, we can construct the correct size $2L$ ground state to linear order in $J/h$, and in fact, the circuit we use to do so is precisely based on the analytic unitaries found in the previous subsection. 
The leading order errors that remain are pairs of spin flips at sites $4i$ and $4i+4$ with coefficients of order $(J/h)^2$. 

If we modify our circuit to contain two spin unitaries acting on pairs of spins located four sites away, we can construct the state \emph{correctly} to order $(J/h)^2$. More generally,  if one continues to add unitaries up to distance $2n$, one can faithfully construct the state to order $(J/h)^n$. The strategy is to first remove unwanted terms to bring the state back to the $J/h=0$ state and then to build in the needed terms.
Both distance $n$ and $2n$ gates are required, the former to create the needed $n^{\textrm{th}}$ order terms, and the latter to remove unwanted $n^{\textrm{th}}$ order terms introduced when we add the ancillae.
Furthermore, the effective Hamiltonians that parameterize these unitaries will be exponentially weak in distance, so the overall unitary  will indeed be quasilocal as expected.
Consistent with our  numerical results, starting with a product state rather than the size $L$ ground state  allows us to do better; in particular, we will only need distance $n$ gates in order to be correct to order $(J/h)^n$ since there are no unwanted terms to remove. We emphasize that this intuition  is only true for sufficiently small $J/h$.

In the ferromagnetic phase things are a bit more subtle. We will denote the ground state which is ``connected'' to the symmetric cat state at $h=0$ as $|\psi_0^L\rangle$ (even parity) and the analogous state which is connected to the anti-symmetric cat state as $|\psi_1^L\rangle$ (odd parity). 
The energy splitting between these states will scale like $(h/J)^L$. 
Were we to only care about energy errors, we might reasonably consider any linear combination of $|\psi_0^{2L}\rangle$ and $|\psi_1^{2L}\rangle$ to be our target state. 
However, we know that one cannot turn a product state into a cat state or vice versa with a circuit of sub-extensive depth. 
As previously discussed, at $h/J=0$ one can go from a cat state input to a cat state output by inserting the ancillae in the $|\up\rangle$ state and using CNOT gates for the bottom layer of the circuit. 
Thus, if our starting state is $|\psi_0^L\rangle$, then our target state will be $|\psi_0^{2L}\rangle$. 
We could also consider starting with either a product state or the symmetry breaking linear combination $(|\psi_0^L\rangle+|\psi_1^L\rangle)/\sqrt{2}$ and building towards $(|\psi_0^{2L}\rangle+|\psi_1^{2L}\rangle)/\sqrt{2}$; in this case the circuit at $h/J=0$ is the identity. 
If we insist that the target is $|\psi_0^{2L}\rangle$ and begin with a product state input it is impossible to be correct to even $0^{\textrm{th}}$ order.

To this end, with the standard circuit (Fig.~\ref{fig:CircuitDiagrams}a), one can prepare $|\psi_0^{2L}\rangle$ correctly only to $\mathcal{O}(1)$, whereas for the symmetry broken or product inputs we can prepare $(|\psi_0^{2L}\rangle+|\psi_1^{2L}\rangle)/\sqrt{2}$ correctly to order $h/J$.
The leading order error in the former case comes in the form of double spin flips on each of the two product states that make up the cat state. 
Using a three site unitary with the third spin acting as a control would let us correct this error and prepare the true ground state correctly to order $h/J$. 
Similarly, blocks of $2n+1$ sites will allow us to correctly prepare the state to order $(h/J)^n$ by eliminating correlated $2n$ spin-flip errors. 
In contrast with the paramagnetic phase, where the long distance unitaries only needed to act on two sites, here the gates need to act on all of the sites within a block. 

When targeting the symmetry broken state, the standard circuit  will give us the correct answer to order $h/J$ for both the product state and symmetry broken inputs. 
However, to correct higher order errors using $(|\psi_0^L\rangle+|\psi_1^L\rangle)/\sqrt{2}$ as an input requires $2n+1$-site unitaries to obtain the correct result at order $(h/J)^{n+1}$; the analogous situation with a  product state input requires $n$ site unitaries to be correct to order $(h/J)^n$. 
In all cases, the strengths of the required unitaries fall off exponentially with the diameter of the block, and the overall circuit is once again quasi-local.

\section{Discussion and Conclusion}
\label{sec:discussion}

In this work, we have developed a numerical implementation of the $s$-source algorithm for finding approximate ground states of local Hamiltonians~\cite{Swingle:2014qpa}.
We approximate the lattice-doubling unitary of the $s$-source algorithm as an efficiently contractable tensor network, which we in turn variationally optimize to minimize the energy of the doubled-lattice ground state. 
To ensure our tensor network is efficiently contractable, we construct it from local rather than quasi-local components, although this decreases the accuracy of the approximation.
We benchmark the resulting numerical algorithm on several 1D spin chain models, and find that the $s$-source construction works particularly well at scale-invariant critical points.
We ascribe this to the fact that ancillae insertion doubles the length scale of all correlations in the input state, much in the same spirit as MERA.
In addition, to gain some analytic intuition, we computed the  scaling of the wavefunction errors deep in each phase of the TFIM, and determined how the $s$-source circuit could be modified to correct these errors. 
These corrections are consistent with the expectation  that performing  $s$-source  with a quasi-local unitary should permit the exact construction of the doubled-lattice ground state.

Our work suggests several interesting directions for future study. 
First, one could use multi-layer $s$-source as a numerical method to  extract information about renormalization group flow.
When creating a multilayer $s$-source circuit, one obtains a sequence of unitaries $U_1$, $U_2$, ..., $U_n$ that each double the size of the system.
By parameterizing how this sequence of unitaries changes, it should be possible to follow the renormalization group flow and to extract quantities such as the operator dimension. 
In a similar vein, since the Hamiltonian parameters also flow, one might expect that the ideal input state to build $\ket{\psi^{2L}(g)}$ would not be $\ket{\psi^{L}(g)}$ but rather $\ket{\psi^{L}(g^\prime)}$, where the Hamiltonian parameter $g^\prime$ at length $L$ flows to $g$ at length $2L$. Allowing for this may significantly improve the performance of $s$-source away form criticality.

While we do not foresee $s$-source outperforming established methods like DMRG in determining 1D ground states, it may be useful for constructing ground states in 2D where existing methods leave more room for improvement.
Furthermore, the algorithm could be naturally adapted as an experimental method for preparing ground states. 
%
%
In fact, the ability to interleave ancillae has recently become possible in Rydberg optical tweezer arrays~\cite{Endres1024, Barredo1021, PhysRevLett.121.123603}.
An experimental implementation of $s$-source would be particularly useful for generating states with long correlation lengths.
Indeed, as we have previously discussed,  strongly-correlated many-body states often require deep quantum circuits in order to be built from product states.
Absent error correction, deep circuits result in low fidelities due to compounding gate errors. 
Thus, the ability of $s$-source to create certain classes of strongly-correlated states with low-depth circuits could provide a significant advantage in the NISQ era \cite{Preskill2018quantumcomputingin}.
Finally, although the numerical implementation we explore here is  variational, the $s$-source formalism provides a compelling connection to a non-variational ground-state construction which merits future exploration.

\emph{Acknowledgments}---We acknowledge the insights of and discussions with  C. Laumann, T. Schuster, F. Machado, K. Akkaravarawong, J. Motruk, W.-W. Ho, G. Evenbly, and I. Cirac. 
We are particularly grateful to M. Zaletel for suggesting the current version of our multilayer optimization approach and R. Mong for suggesting the idea of using $s$-source as a method for numerically computing renormalization group flows.
JM is grateful to B. Swingle for collaboration on the $s$-sourcery program.
DMRG calculations were performed using the ITensor Library\cite{ITensor}.
This work was supported by the U.S. Department of Energy, Office of Science, Office of Advanced Scientific Computing
Research, Quantum Algorithm Teams Program, the U.S. Department of Energy, Office of Basic Energy Sciences, Division of Materials Sciences and Engineering under Award DE-SC0019241, and the U.S. Department of Energy under cooperative research agreement DE-SC0009919. 
SG acknowledges support from the Israel Science Foundation, Grant No. 1686/18.
CO was supported by the Department of Defense through the National Defense Science \& Engineering Graduate Fellowship program.


\bibliographystyle{apsrev4-1}
\bibliography{ssource}

\appendix
\section{Circuit optimization}
\label{A:optimization}

Finding a quantum circuit that prepares a minimum energy eigenstate is a challenging problem \cite{arad_quantum_2010, markov_simulating_2008}. 
In the generic case, we cannot deterministically find the optimal circuit or even verify that a given circuit does minimize the energy.
We can, however, perform a variational search by starting with some circuit and then iteratively updating component tensors in order to continually lower the energy, as is common practice for MERA. 

Within a superlayer of the $s$-source circuit, we optimize each local unitary $u$ while holding all others fixed.
We then iterate this procedure for all $u$ in the superlayer until the energy has converged.
To update a single unitary, we utilize the optimization strategy outlined in \cite{PhysRevB.79.144108}. 
To summarize, we pretend that the function we are trying to maximize, $f(u)=-\langle H\rangle$, is a linear function of $u$. 
Of course it is actually quadratic, as $u^\dagger$ is present in the dual circuit.
If we treat $u^\dagger$ as constant, however, we can write $f(u)=\tr(W^\dagger u)$, where the environment $W^\dagger$ of $u$ is found by contracting the tensor network formed by removing $u$ from $-\langle H\rangle$'s tensor network. 
If $W$ has singular value decomposition $W=XYZ^\dagger$, it follows that a linear function $f(u)$ achieves its maximum at $u=XZ^\dagger$. 

For a nonlinear function of $u$, one should in principle update $u$ multiple times until the energy converges.
In practice, we only update each tensor $u$ once during a full sweep of the superlayer; we have empirically found that this leads to a lower energy for a fixed total number of updates.
We also tested an alternative optimization strategy referred to as ``Linearization II" in \cite{0707.1454}, but this required using ~10 updates per unitary per superlayer sweep for numerical stability, as well as tuning additional hyperparameters. 
We did not see any benefit of this approach for fixed computational cost.

As one might expect, this update procedure generally only finds a local energy minimum, not a global one. 
In order to get the global minimum we perform this optimization many times over circuits initialized with Haar-random unitaries. 
The cost of simulation is linear in the number of samples, which can be large, so we note a few tricks tricks that will improve either the speed or performance of optimization (although sampling over initializations is embarrassingly parallel). 

First, we note that only a subset of the terms in the Hamiltonian will be within the light cone of a given $u$, so we only need to minimize the partial energy containing those terms when we update $u$. 
Crucially, the number of terms that contribute is \emph{constant} as a function of system size, whereas the number of terms in the full Hamiltonian scales as $L$. 
We also note that each optimization step does not, generally, decrease the energy.
For a quadratic function of a given $u$ this procedure will actually maximize the absolute value of the function.
In order to avoid this complication, we alter the spectrum of the (partial) energy we are minimizing to be negative-definite by shifting the partial hamiltonian by an appropriate multiple of the identity \cite{Evenbly2013}: $\hat{h}\rightarrow \hat{h}-\alpha I$, where $\alpha$ is the maximum eigenvalue of the partial hamiltonian $\hat{h}$. 
In practice we find that we only need to do this for a few sweeps before all partial energies are negative, at which point we turn the shift off as it seems to slow down convergence (this suggests the possibility that shifting the spectrum \textit{up} could speed up convergence as long as we are careful to keep things negative-definite).

We now describe some heuristics for efficient contraction of the next-nearest-neighbor $s=1$ $s$-source circuit. 
Suppose we want to evaluate the expectation value of a term in the hamiltonian:
\begin{equation}
\langle\hat{O}\rangle=\langle\psi^L| U^\dagger_1\ldots U^\dagger_{n}\hat{O}U_{n}\ldots U_1|\psi^L\rangle
\end{equation}
where $U_i$ is the $i$th superlayer of a multilayer $s$-source circuit.
We start with the operator $\hat{O}$ (which is defined on the $2^n L$ site lattice) and then, in the language of MERA, act upon it with the ascending superoperator \cite{PhysRevB.79.144108} (in other words, we conjugate by the innermost superlayer of the circuit). 
If $\hat{O}$ was supported on at most 6 adjacent sites, the ascended version of $O$ will be supported on either 4 or 6 adjacent sites on the $2^{n-1}L$ spin lattice. 
This is because the causal cone extends by at most 6 sites, and contracting with the ancillae halves the support of the operator at the end. 

We emphasize that we can ignore contraction with any gate in $U_n$ outside of $\hat{O}$'s causal cone, since it will contract with its inverse in $U^\dagger_n$ to form an identity \cite{PhysRevB.79.144108}. 
Therefore, there is no computational advantage to starting with a block of fewer than 4 sites (or 6 sites if the block would be ascended to a 6 site block), so if we want to evaluate the sum of expectation values of many operators, we should group them into blocks of operators living on either 4 or 6 adjacent sites. 
Doing so allows us to do a single contraction to find the sum of the expectation values of several adjacent local observables instead of multiple contractions to find the expectation value of each term separately.
This reduces the time it takes to evaluate the expectation value of the hamiltonian considerably.
Repeatedly applying ascending superoperators, we eventually obtain an operator defined on the $L$ site lattice, which when contracted with the MPS and its dual will give the desired expectation value. 

It is advantageous to cache various partial contractions of the MPS portion of the tensor network (i.e.~$\bra{\psi^L}$ and $\ket{\psi^L}$).
In particular, at the final step of evaluating an expectation value, we will contract the ascended $\hat{O}$ with 4 or 6 pairs of adjacent physical indices of the MPSs, with all other MPS indices already contracted. 
By storing all contractions of the MPSs with 4 or 6 pairs of adjacent dangling bonds, we can avoid repeating this costly computation.
The evaluation of the $W^\dagger$s needed to optimize the circuit is done in much the same way, simply omitting the contraction with the specific $u$ that is to be updated.

Energy minimization appears to take on the order of 1000 sweeps for the models we tested, with that number growing slightly with system size. 
This also varies from run to run; sometimes it might take 1000 sweeps, and sometimes it may take 10000. 
For most of the figures in this paper starting with an initial state of 32 spins and creating an $s$-source state of 64 spins we ran  $\sim 1000$ initializations with 1000 sweeps each and took the best energy among them. 
For smaller systems, e.g. 8 to 16, we performed $<100$ initializations. 

We note that for this work we were particularly interested in characterizing the error of the $s$-source algorithm, and as such we needed to find the global minimum as reliably as possible. 
For some other applications, one may be perfectly content to have, say, twice the minimum error, in which case it is not as necessary to run so many randomly initialized optimizations. 
In this case one can often do pretty well by starting with a good guess for the initial circuit, adding some noise, and optimizing just a few initial states. 
For the TFIM, a good guess may be the leading order analytic solution that we discuss in Sec.~\ref{sec:large-h}, where noise is added by multiplying each unitary by another random unitary close to the identity.
Here, we make two notes. First, with fewer parallel optimizations it is more important to do more sweeps for each one (several thousand rather than one thousand, say). 
Secondly, it is important to make sure that the initial condition of the circuit is not entirely real, as updating a real valued circuit will keep the circuit in the real manifold. 

We note that further improvements are likely possible. 
It seems, for example, that it should be possible to reduce the average number of required sweeps by monitoring for convergence.  
However, checking for convergence can be quite deceptive here; one typically sees plateaus where the energy appears to converge, and then sudden jumps down to new local minima. 
A more careful analysis may reveal an effective way to anticipate whether or not further sweeps will result in an improved energy. 
In our experience, the energy would sometimes continue to improve beyond 1000 sweeps, but it was more efficient to sample more initial conditions than execute more sweeps per sample.

Finally, the cost of contraction scales roughly exponentially in the width of the circuit's causal cone.
In practice, this might motivate the use of the simplified $s$-source circuit comprised of only the nearest-neighbor gates.
We analyzed this circuit in Sec.~\ref{sec:results} and found that the energy error was qualitatively similar to that of the circuit containing both nearest-neighbor and next-nearest-neighbor gates, and was quantitatively not much worse at the critical point.

\section{Derivation of analytic unitaries for large magnetic fields}
\label{A:large-h}
Here we derive the analytic expressions for the s-source unitaries of the TFIM in the limit $h\gg J$, previously described in Sec.~\ref{sec:large-h}. 
In this regime, we can simultaneously make the adiabatic evolution time $T$ long enough to be adiabatic, but short enough that we can do a Trotter expansion. 
The former condition requires, for $h\gg J$, $hT \gg 1$. 
In the interaction picture that we will consider shortly, the Trotter expansion requires $JT\ll 1$. 
We consider a single term in the Trotter expansion of Eq.~\ref{eq:adiabaticU} and thus reduce the problem to considering two spins that are initially in a field of strength $h$ in the $x$ direction and then turning on an interaction of strength $J$ in the $z$ direction. 
If we slowly turn on the interaction over a time $T$, then we have
\begin{equation}
H(t)=H_0+\frac{t}{T} H_1
\end{equation}
with
\begin{equation}
H_0=-h(\sigma_x^1+\sigma_x^2)
\end{equation}
\begin{equation}
H_1=-J\sigma_z^1\sigma_z^2.
\end{equation}
Then the adiabatic unitary associated with moving from $H_0$ to $H_0+H_1$ is
\begin{equation}
U=\mathcal{T}\exp{-i\int_0^T H(t)dt}.
\end{equation}

It is helpful for us to move to the interaction picture before proceeding. Doing so gives us the interaction picture unitary
\begin{widetext}
\begin{equation}
U_I=\mathcal{T}\exp{-i\int_0^T \frac{t}{T}e^{iH_0 t}H_1e^{-iH_0 t}dt}\approx 1 -i\int_0^T \frac{t}{T}e^{iH_0 t}H_1e^{-iH_0 t}dt.
\end{equation}
Upon integrating and discarding higher order terms, we get

\begin{equation}
U_I  = 1 + i\left(\frac{JT}{4}(\sigma_z^1\sigma_z^2+\sigma_y^1\sigma_y^2)+\frac{J}{8h}e^{-ih(\sigma_x^1+\sigma_x^2)T}(\sigma_z^1\sigma_y^2+\sigma_y^1\sigma_z^2)e^{ih(\sigma_x^1+\sigma_x^2)T}\right).
\end{equation}
Moving back to the Schrodinger picture and continuing to work to leading order,
\begin{equation}
\begin{split}
U & = e^{-iH_0 T}U_I = \left(e^{-iH_0 T}U_I e^{iH_0 T}\right)e^{-iH_0 T}\\
& = 1 + i\left(hT(\sigma_x^1+\sigma_x^2)+\frac{JT}{4}(\sigma_z^1\sigma_z^2+\sigma_y^1\sigma_y^2)+\frac{J}{8h}(\sigma_z^1\sigma_y^2+\sigma_y^1\sigma_z^2)\right)\\
& = \exp i\left(hT(\sigma_x^1+\sigma_x^2)+\frac{JT}{4}(\sigma_z^1\sigma_z^2+\sigma_y^1\sigma_y^2)+\frac{J}{8h}(\sigma_z^1\sigma_y^2+\sigma_y^1\sigma_z^2)\right) \\
& = \exp( -iH_\mathrm{eff})
\end{split}
\end{equation}
which corresponds to an effective Hamiltonian
\begin{equation}
H_\mathrm{eff} =-hT(\sigma_x^1+\sigma_x^2)-\frac{JT}{4}(\sigma_z^1\sigma_z^2+\sigma_y^1\sigma_y^2)-\frac{J}{8h}(\sigma_z^1\sigma_y^2+\sigma_y^1\sigma_z^2).
\end{equation}
\end{widetext}

We have not as of yet specified a value for $T$, so its presence in our effective Hamiltonian may appear, at first glance, to be troubling. 
However, we expect from the adiabatic theorem that, as long as the assumptions are met, there should be no strong $T$ dependence. 
Indeed, one can explicitly verify that in the limit $T\gg J/h^2$ the effect of the $T$ dependant terms is a phase shift. 
Dropping them, we end up with a particularly simple form for $H_\mathrm{eff}$:
\begin{equation}
H_\mathrm{eff} =-\frac{J}{8h}(\sigma_z^1\sigma_y^2+\sigma_y^1\sigma_z^2).
\end{equation}

The ``off" unitaries, on the other hand, are given by $\exp{(iH_\mathrm{eff})}$. 
We can see this by considering running the process backwards in time. 
This is, of course, just the ``turn on" problem we just solved. 
There is one additional complication: these unitaries are acting not on the ground state, but on the first layer of the circuit. 
However, corrections due to the non-commutation of the the layers will come in at a higher order, and since we are only working to first order anyway we can simply ignore them.

We can repeat this analysis for a mixed coupling and field Ising model. Here we have hamiltonians
\begin{widetext}
\begin{equation}
H_0=-h_x(\sigma_x^1+\sigma_x^2)-h_z(\sigma_z^1+\sigma_z^2)
\end{equation}
and
\begin{equation}
H_1=-J_x\sigma_x^1\sigma_x^2-J_z\sigma_z^1\sigma_z^2.
\end{equation}
If we define $h=\sqrt{h_x^2+h_z^2}$, $\tan\eta = h_z/h_x$, and go through the same steps, we find that to order $J/h$
\begin{equation}
\begin{split}
H_\mathrm{eff,XX}= & -\frac{J_xT}{32}\left(9+4\cos{2\eta}+3\cos{4\eta}\right)\sigma_x^1\sigma_x^2-\frac{J_xT}{4}\sin^2\eta\,\sigma_y^1\sigma_y^2-\frac{3J_xT}{16}\sin^2{2\eta}\,\sigma_z^1\sigma_z^2 \\
& -\frac{J_xT}{32}\left(2\sin{2\eta}+3\sin{4\eta}\right)(\sigma_x^1\sigma_z^2+\sigma_z^1\sigma_x^2) + \frac{J_x}{32h}\left(7\sin{\eta}+3\sin{3\eta}\right)(\sigma_x^1\sigma_y^2+\sigma_y^1\sigma_x^2) \\
& + \frac{3J_x}{8h}\cos\eta\,\sin^2\eta\,(\sigma_y^1\sigma_z^2+\sigma_z^1\sigma_y^2)
\end{split}
\end{equation}
\begin{equation}
\begin{split}
H_\mathrm{eff,ZZ}= & -\frac{3J_zT}{16}\sin^2{2\eta}\,\sigma_x^1\sigma_x^2-\frac{J_zT}{4}\cos^2\eta\,\sigma_y^1\sigma_y^2-\frac{J_zT}{32}\left(9-4\cos{2\eta}+3\cos{4\eta}\right)\sigma_z^1\sigma_z^2 \\
& -\frac{J_zT}{32}\left(2\sin{2\eta}-3\sin{4\eta}\right)(\sigma_x^1\sigma_z^2+\sigma_z^1\sigma_x^2) -\frac{3J_z}{8h}\cos^2\eta\,\sin\eta\,(\sigma_x^1\sigma_y^2+\sigma_y^1\sigma_x^2) \\
& -\frac{J_z}{32h}\left(7\cos{\eta}-3\cos{3\eta}\right)(\sigma_y^1\sigma_z^2+\sigma_z^1\sigma_y^2),
\end{split}
\end{equation}
with an overall effective hamiltonian $H_\mathrm{eff}=H_0T+H_\mathrm{eff,XX}+H_\mathrm{eff,ZZ}$. If we again drop the $T$ dependent terms, we get
\begin{equation}
\begin{split}
H_\mathrm{eff}= & \left(\frac{J_x}{32h}\left(7\sin{\eta}+3\sin{3\eta}\right) -\frac{3J_z}{8h}\cos^2\eta\,\sin\eta\right)(\sigma_x^1\sigma_y^2+\sigma_y^1\sigma_x^2) \\
& + \left(\frac{3J_x}{8h}\cos\eta\,\sin^2\eta-\frac{J_z}{32h}\left(7\cos{\eta}-3\cos{3\eta}\right)\right)(\sigma_y^1\sigma_z^2+\sigma_z^1\sigma_y^2).
\end{split}
\end{equation}
\end{widetext}


\end{document}